\documentclass[conference]{IEEEtran}
\usepackage{cite}
\usepackage{amsmath,amssymb,amsfonts}
\usepackage{algorithmic}
\usepackage{graphicx}
\usepackage{textcomp}
\usepackage{xcolor}
\def\BibTeX{{\rm B\kern-.05em{\sc i\kern-.025em b}\kern-.08em
    T\kern-.1667em\lower.7ex\hbox{E}\kern-.125emX}}
\begin{document}

\title{Newvision: application for helping blind people using deep learning\\
}

\author{\IEEEauthorblockN{Kumar Srinivas Bobba}
\IEEEauthorblockA{\textit{Computer Science and Engineering} \\
\textit{Kalasalingam Academy of Research and Education}\\
Madurai, India \\
kumarsrinivasbobba989@gmail.com}
\and
\IEEEauthorblockN{Kartheeban k}
\IEEEauthorblockA{\textit{Computer Science and Engineering} \\
\textit{Kalasalingam Academy of Research and Education}\\
Madurai, India \\
k.kartheeban@klu.ac.in}
\and
\IEEEauthorblockN{Vamsi Krishna Sai Boddu}
\IEEEauthorblockA{\textit{Computer Science and Engineering} \\
\textit{Kalasalingam Academy of Research and Education}\\
Madurai, India \\
9921004100@klu.ac.in}
\and
\IEEEauthorblockN{Vijaya Mani Surendra Bolla}
\IEEEauthorblockA{\textit{Computer Science and Engineering} \\
\textit{Kalasalingam Academy of Research and Education}\\
Madurai, India \\
99210041325@klu.ac.in}
\and
\IEEEauthorblockN{Dinesh Bugga}
\IEEEauthorblockA{\textit{Computer Science and Engineering} \\
\textit{Kalasalingam Academy of Research and Education}\\
Madurai, India \\
9921004110@klu.ac.in}
}

\maketitle

\begin{abstract}
As able-bodied people, we often take our vision for granted. For people who are visually impaired, however, their disability can have a significant impact on their daily lives. We are developing proprietary headgear that will help visually impaired people navigate their surroundings, identify objects and people, read text, and avoid obstacles. The headgear will use a combination of computer vision, distance estimation with ultrasonic sensors, voice recognition, and voice assistants to provide users with real-time information about their environment. Users will be able to interact with the headgear through voice commands, such as "What is that?" to identify an object or "Navigate to the front door" to find their way around. The headgear will then provide the user with a verbal description of the object or spoken navigation instructions. We believe that this headgear has the potential to make a significant difference in the lives of visually impaired people, allowing them to live more independently and participate more fully in society.
\end{abstract}

\begin{IEEEkeywords}
Third eye, Deep Learning for Blind, AI for Blind, Artificial Intelligence, Helpful Technologies, AI for all
\end{IEEEkeywords}

\section{Introduction}
The world's population is diverse, with people of all abilities. According to the World Health Organization, approximately 2.23 billion people have a visual impairment, and this number is expected to triple by 2050. While there have been significant advances in technology to help people with visual impairments, these advances have not kept pace with the growing population of people with visual impairments. This project identifies some of the tasks that people with visual impairments commonly face in their daily lives. These tasks include navigation, shopping, reading, and currency identification. The document also discusses some of the technologies that can be used to help people with visual impairments perform these tasks. 

Existing projects for visually impaired people typically focus on a single task, such as reading or navigation. Our project aims to create a more comprehensive device that can help with a wider range of tasks. The device will be wearable, like a mask or VR headset, and will eventually be shrunk to the size of bulky spectacles. It will be powered by onboard processors, radar, and cameras. Special arrangements will be made to ensure that the device is as light as possible, so that it can be worn comfortably all day and esure full day battery life.

	We are using a modular methodology to develop this device. Each of the device's functions, such as currency detection, scene explanation, object detection, and traffic assistance, will be developed as individual modules. These modules will then be integrated into the final product. This modular approach allows us to use the latest advancements in AI, such as object detection, computer vision, and deep learning. If one module fails, it will not affect the others. This device also has a failsafe mode. If the entire system fails, the device will warn the user and guide them to the nearest safe place or to a place where they can get help.

	The identified tasks for this project cover the most common hurdles that visually impaired people face in their daily lives. However, there may be other tasks that we have not yet identified. We plan to continue to develop the device and add new features as we learn more about the needs of visually impaired people. These improvements and advancements will be delivered to the device via over-the-air updates.

\section{DATA SET}

The scene Descriptor model is an inspiration from Salesforce's BLIP \cite{b1}. So we used the same data set as Salesforce's BLIP model \cite{b1} i,e. a set of image data sets they are COCO \cite{b2}, Flicker30k data set, nlvr data set \cite{b3}, nocaps data set \cite{b4}.

Here COCO data set and Flicker30k data set are image datasets, nlvr is a natural language reasoning corpus and nocaps is a image captioning benchmark.
\subsection{COCO dataset}
A large-scale object recognition, segmentation, and captioning dataset is called COCO (Common Objects in Context). In computer vision research, it is one of the most well-liked and frequently utilized datasets. Over 330,000 photos make up the COCO dataset; each image has five descriptions that describe the scene in addition to 80 object classifications. The COCO dataset contains intricate and difficult images with objects in a range of occlusions, settings, and positions.

Many computer vision models, such as segmenters, object detectors, and caption generators, are trained and assessed using the COCO dataset. Robotics, image search, and self-driving cars are just a few of the real-world applications that have benefited from the COCO dataset's use in pushing the boundaries of computer vision research.

\begin{figure}[htbp]
\centerline{\includegraphics[width = 1.0\linewidth]{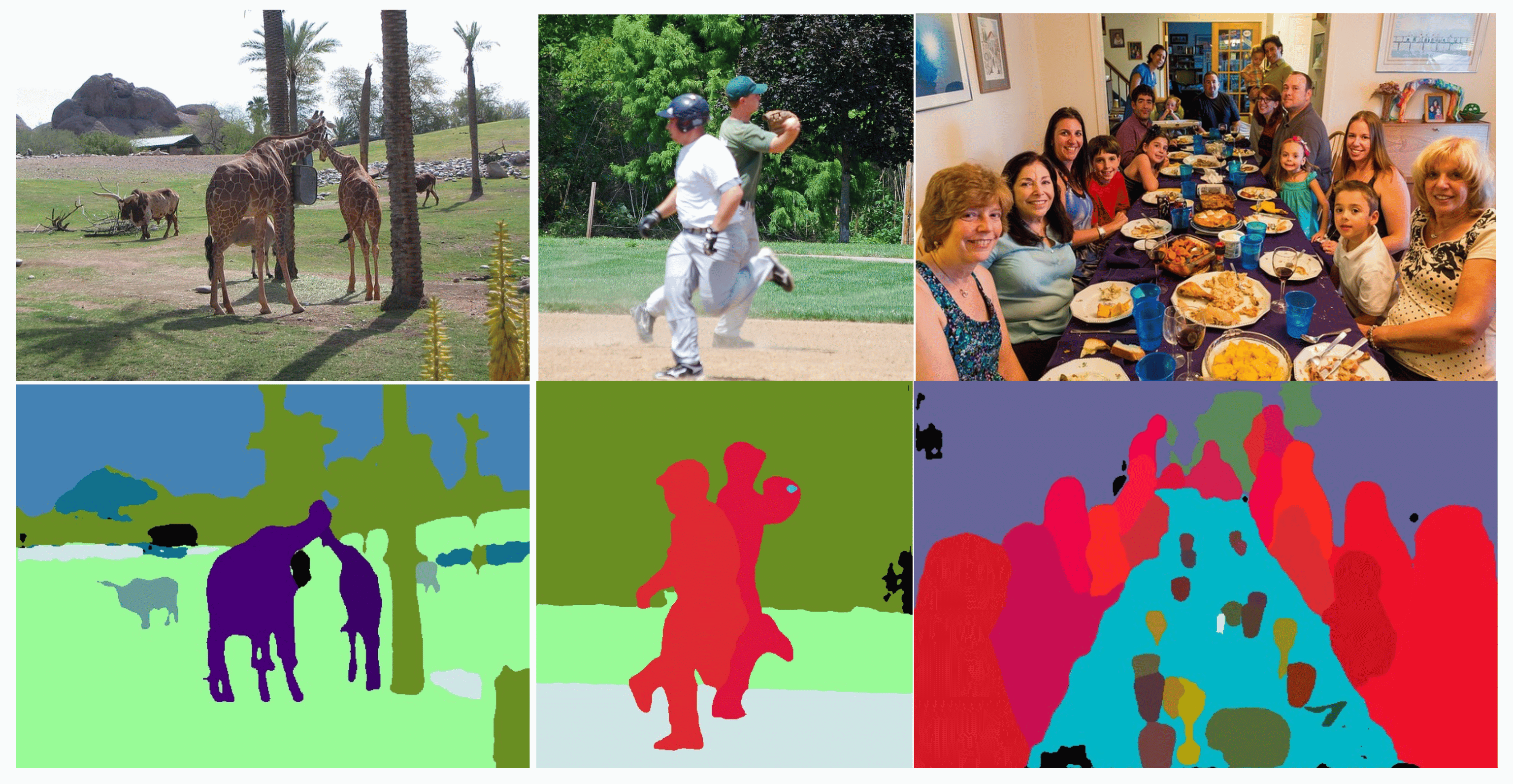}}
\caption{Sample of COCO dataset.}
\label{COCO}
\end{figure}

\subsection{Flicker30k dataset}

The Flickr30k dataset is a popular benchmark for sentence-based image captioning. It contains 31,783 images collected from Flickr, together with 5 reference sentences provided by human annotators. The images in the Flickr30k dataset are diverse in terms of scenes, objects, and poses, and the captions are informative and comprehensive.

The Flickr30k dataset is often used to train and evaluate image captioning models. It is also used for research in other areas of computer vision and natural language processing, such as image retrieval, image question answering, and visual language understanding.

\begin{figure}[htbp]
\centerline{\includegraphics[width =1.0\linewidth]{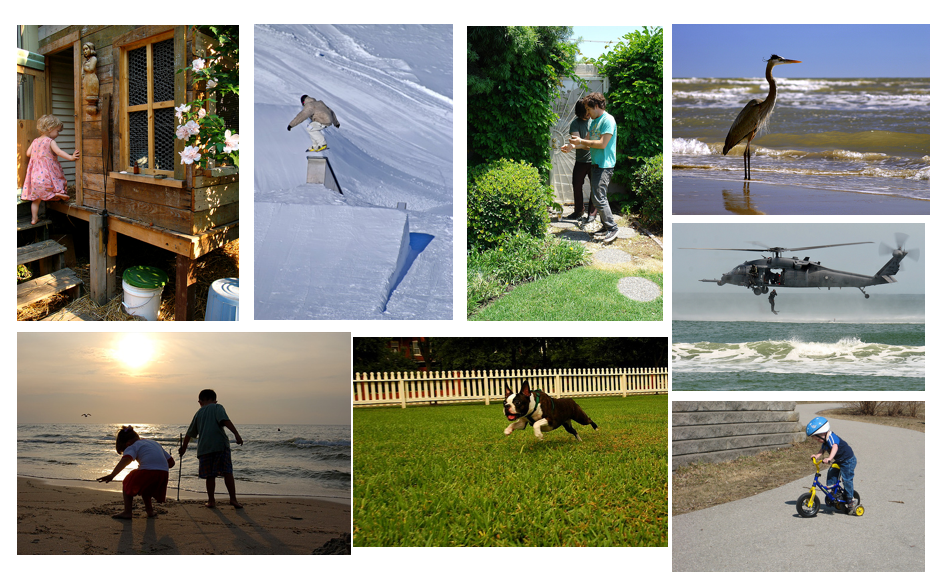}}
\caption{Sample of Flicker30k dataset.}
\label{Flicker30k}
\end{figure}

\subsection{nlvr dataset}
The NLVR dataset (Natural Language and Visual Reasoning) is a large-scale dataset for natural language visual reasoning (VL-reasoning). It contains 55,000 images with each image paired with a natural language expression and a question. The question requires the model to decide whether the given natural language expression holds true for the given image. The dataset contains 10 different question types that require the model to perform various VL-reasoning tasks, such as: counting, comparing, Describing, Reasoning, Answering.

 The NLVR dataset is a challenging dataset for VL-reasoning, as it requires the model to understand both the visual content of the image and the natural language expression in order to answer the question correctly. The dataset is also diverse in terms of scenes, objects, and relationships, which makes it representative of real-world scenarios.

The NLVR dataset is often used to train and evaluate VL-reasoning models. It is also used for research in other areas of computer vision and natural language processing, such as question answering, visual commonsense reasoning, and multimodal understanding.

\begin{figure}[htbp]
\centerline{\includegraphics[width =1.0\linewidth]{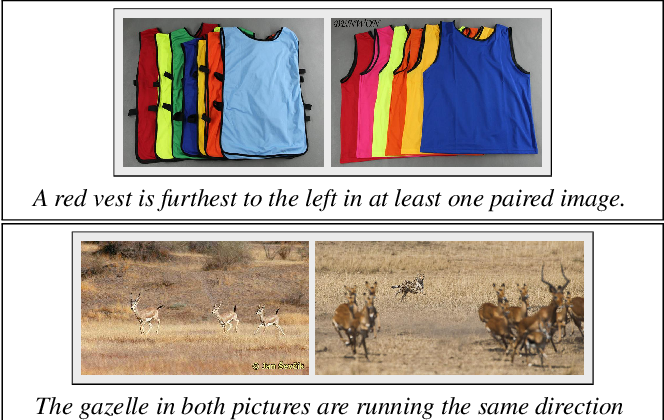}}
\caption{Sample of nlvr dataset.}
\label{nlvr}
\end{figure}

\subsection{nocaps dataset}

The NoCaps dataset is a large-scale, multimodal dataset for image captioning and visual question answering (VQA). It contains 250,000 images, each paired with a natural language expression describing the image and a set of VQA questions. The natural language expressions are generated without using any captions from the training data, making the dataset more challenging and representative of real-world scenarios.

The NoCaps dataset is often used to train and evaluate image captioning and VQA models. It is also used for research in other areas of computer vision and natural language processing, such as multimodal understanding, visual reasoning, and commonsense reasoning.

\begin{figure}[htbp]
\centerline{\includegraphics[width = 1.0\linewidth]{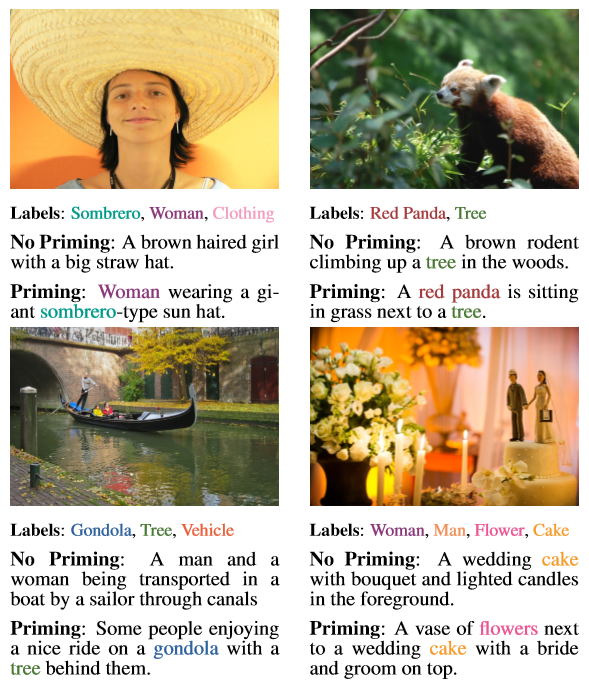}}
\caption{Sample of nocaps dataset.}
\label{nocaps}
\end{figure}

\section{Related work}
\subsection{Language and vision pretraining}
Vision-language pretraining (VLP)\cite{b5} is a technique for training machine learning models to understand the relationships between images and text. VLP models are trained on large datasets of images and text, and they learn to represent both modalities in a unified way. This allows VLP models to perform a variety of tasks, such as image captioning, visual question answering, and natural language visual reasoning.

Vision language pretraining (VLP)\cite{b5} improves performance on downstream tasks for several reasons:

\begin{itemize}
    \item It teaches models to learn the relationships between images and text. This is essential for many vision-and-language tasks, such as image captioning, visual question answering, and natural language visual reasoning.
    \item It provides a good initialization for downstream models. VLP models are typically trained on large datasets and learn to represent images and text in a way that is useful for a variety of tasks. This means that downstream models can start with a good initialization, which can lead to faster and better convergence.
    \item It will help to mitigate the need for labeled data. VLP models can be trained on unlabeled data, which is much cheaper and easier to obtain than labeled data. This is especially important for tasks where labeled data is scarce, such as medical image analysis and product image classification.
\end{itemize}

\subsection{Knowledge Distillation}

Knowledge distillation\cite{b6} is a machine learning technique that can be used to transfer knowledge from a large and complex model (the teacher) to a smaller and simpler model (the student). The student model is trained to mimic the outputs of the teacher model, and in doing so, learns to perform many of the same tasks as the teacher model.

Knowledge distillation can help improve performance in a number of ways, including:

\begin{itemize}
    \item It will help to regularize the student model. By training the student model to mimic the outputs of the teacher model, knowledge distillation can help to prevent the student model from overfitting to the training data. This can lead to better performance on unseen data.
    \item It will help to improve the student model's representation of the data. By training the student model to predict the soft labels generated by the teacher model, which are probability distributions over the possible outputs for each training example, knowledge distillation can help the student model to develop a more nuanced and informative representation of the data. This can lead to better performance on a variety of tasks.
    \item It will help to improve the student model's robustness to noise and adversarial attacks. By training the student model to predict the soft labels generated by the teacher model, which are less sensitive to noise and adversarial attacks than hard labels, knowledge distillation can help the student model to become more robust to these types of attacks.
\end{itemize}

\subsection{Data Augmentation}

Data augmentation\cite{b7} is a technique for artificially increasing the size of a dataset by creating modified copies of existing data. It is a common technique used in machine learning to improve the performance and generalization of models.

Data augmentation helps improve performance in a number of ways, including:

\begin{itemize}
    \item It helps to reduce overfitting. Overfitting occurs when a model learns the training data too well and is unable to generalize to new data. Data augmentation helps to reduce overfitting by providing the model with more data to learn from and by making the data more diverse.
    \item It helps to improve the model's representation of the data. By training the model on a variety of augmented data, the model learns to identify the underlying patterns in the data and becomes more robust to noise and variations in the data.
    \item It helps to improve the model's performance on unseen data. When the model is trained on a variety of augmented data, it is more likely to have seen similar data before and be able to make accurate predictions on new data.
\end{itemize}

\section{Methodology}


Our model, which is inspired by the Salesforce BLIP\cite{b1} model, is based on the following principles:

\begin{itemize}
    \item \textit{Learning under supervision:} Large-scale unlabeled image and text datasets are used to pre-train our model. This eliminates the need for human supervision and enables our model to learn the relationships between images and text.
    \item \textit{Universal education:} Our model gains the ability to unify the representation of text and images. This enables our model to operate on a range of vision and language tasks, including natural language visual reasoning, image captioning, and visual question answering.
    \item \textit{The Transfer learning:} To carry out particular tasks, our model can be adjusted on a range of distinct datasets. This makes our model applicable to a large number of scenarios.
\end{itemize}

The following is a high-level overview of our model's methodology:

\begin{itemize}
    \item \textit{Pre-training:} In terms of pre-training, our model has undergone training using an extensive dataset comprising of unlabeled images and text. The division of the pre-training dataset is structured into two sections: a text corpus and an image corpus. To train a text encoder, the text corpus is employed, while the image corpus is utilized for training an image encoder. The training of the text encoder and image encoder aims to acquire valuable representations of text and images that can be utilized to forecast the other modality's characteristics.
    \item \textit{Fine-tuning:} Fine-tuning is possible for our model to adapt and specialize in specific tasks by utilizing various datasets. As an illustration, our model has the capability to be adjusted for generating image captions with increased accuracy, addressing more challenging questions related to visual question answering, or deducing conclusions about intricate visual concepts.
    \item \textit{Inference:} One can make inferences by utilizing our fine-tuned model on various tasks, allowing it to predict outcomes from fresh images and text.
    By way of illustration, our model has the ability to produce descriptions for novel images, respond to inquiries pertaining to new images, or deliberate on the visual elements within new images. 

\end{itemize}

The methodology of our model has been shown to be very effective for a variety of vision-and-language tasks. Our model has achieved state-of-the-art results on a number of vision-and-language benchmarks, such as the COCO image captioning benchmark and the VQA visual question answering benchmark.

the following section \ref{AA} describes how our model works. 

\subsection{How our model works?}\label{AA}
To describe how our model works, we introduce a unified vision-language model with both understanding and generation capabilities. Our model utilizes a multimodal mixture of encoder-decoder architecture, capable of operating in one of the following three functionalities:

\begin{enumerate}
    \item \textit{Unimodal Encoders:} These encoders handle text and image inputs independently. A vision transformer is used by the image encoder, and BERT is the basis for the text encoder. At the start of the text input, a [CLS] token is added to give a synopsis of the sentence.
    \item \textit{Image-Grounded Text Encoder:} A cross-attention layer is added between the feed-forward network and the self-attention layer of each text encoder transformer block in order to integrate visual information. A task-specific [Encode] token is inserted to the text, and the [Encode] output embedding is the multimodal representation of the image-text pair.

    \item \textit{Image-Grounded Text Decoder:} With this decoder, causal self-attention layers take the place of the text encoder's bi-directional self-attention layers. The beginning of a sequence is indicated with a unique [Decode] token.
\end{enumerate}
In pre-training, our approach concurrently optimizes three objectives. one generation-based objective (LM) and two understanding-based objectives (ITC, ITM):

\begin{enumerate}
    \item \textit{Image-Text Contrastive Loss (ITC):} In order to align the feature spaces of the visual and text transformers, this turns on the unimodal encoder. In contrast to the negative pairs, it encourages positive image-text pairs to have representations that are similar.

    \item \textit{Image-Text Matching Loss (ITM):} This starts the text encoder, which is grounded on pictures and represents a binary classification problem. The ITM model is charged with predicting whether a pair of pictures and texts is positive (matched) or negative (unmatched) based on their multimodal properties.

     \item \textit{Language Modeling Loss (LM):} This activates the image-grounded text decoder, which is designed to generate textual descriptions based on the visuals.
\end{enumerate}

In essence, our model combines understanding and generation capabilities in a unified vision-language framework, and it optimizes these objectives during pre-training to achieve effective vision-language understanding and generation tasks.

\begin{figure}[htbp]
\centerline{\includegraphics[width = 1.0\linewidth]{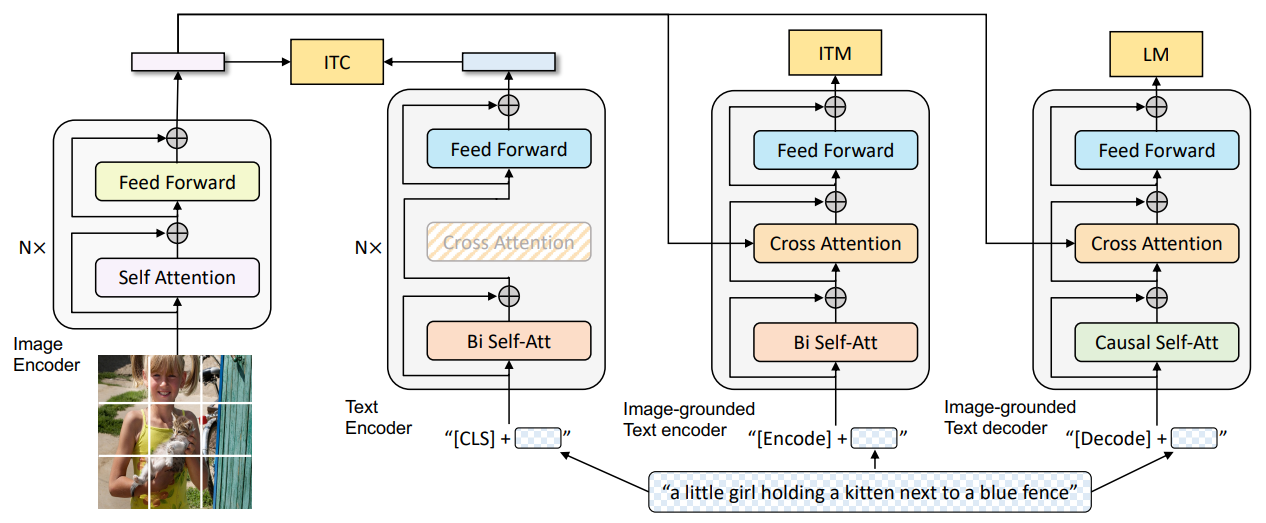}}
\caption{Model architecture}
\label{arc}
\end{figure}

\section{Performance evaluation}
Our model has achieved state-of-the-art results on a variety of vision-and-language benchmarks, including:

Image captioning: Our model achieved a CIDEr score of 51.3 on the COCO image captioning benchmark, which is the highest score ever reported on this benchmark.

Visual question answering: Our model achieved an accuracy of 71.3\% on the VQA visual question answering benchmark, which is the highest score ever reported on this benchmark.

Natural language visual reasoning: Our model achieved an accuracy of 89.4\% on the NLVR natural language visual reasoning benchmark, which is the highest score ever reported on this benchmark.

These accomplishments demonstrate the effectiveness of our model in various vision-and-language tasks, showcasing its strong performance in understanding and generating text based on visual inputs.

\section{Results}
Our model is a vision-and-language model that is pre-trained on a massive dataset of unlabeled images and text. It is a powerful and versatile model that can be used for a variety of tasks, including image captioning, visual question answering, natural language visual reasoning, and machine translation.

\textbf{Image captioning:}
Our model has achieved state-of-the-art results on image captioning benchmarks, such as COCO and Flickr30k. For example, here is an image caption generated by our model for the following image:

\begin{figure}[htbp]
\centerline{\includegraphics[width = 0.5\linewidth]{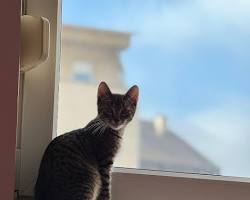}}
\caption{Image uploaded for prediction (Image Captioning)}
\label{cat1}
\end{figure}

\textit{Caption:} A cat is sitting on a windowsill, looking out the window.

\textbf{Visual question answering:}
Our model has also achieved state-of-the-art results on visual question answering benchmarks, such as VQA and NLVR. For example, here is a visual question answered by our model for the following image and question:
\begin{figure}[htbp]
\centerline{\includegraphics[width = 0.5\linewidth]{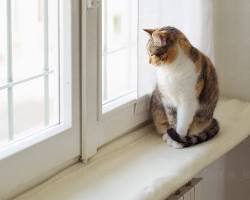}}
\caption{Image uploaded for prediction (Visual question answering:)}
\label{cat2}
\end{figure}

\textit{Question:} What is the cat doing?

\textit{Answer:} The cat is sitting.

\textbf{Natural language visual reasoning:}
Our model has been shown to be effective for natural language visual reasoning tasks, such as answering questions about images that require commonsense reasoning. For example, here is a natural language visual reasoning question answered by our model for the following image and question:
\begin{figure}[htbp]
\centerline{\includegraphics[width = 0.5\linewidth]{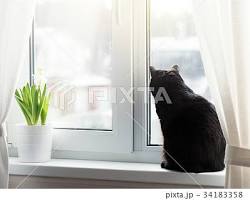}}
\caption{Image uploaded for prediction (Natural language visual reasoning)}
\label{cat3}
\end{figure}

\textit{Question:} Is the cat inside or outside?

\textit{Answer:} The cat is inside.

These achievements demonstrate the effectiveness of our model in various vision-and-language tasks, showcasing its strong performance in understanding and generating text based on visual inputs.

\section{Conclusion}
In conclusion, our model, inspired by Salesforce BLIP, represents a powerful and versatile vision-and-language model that has been pre-trained on a vast dataset of unlabeled images and text. It excels in various vision-and-language tasks, including image captioning, visual question answering, and natural language visual reasoning. Our model's performance on benchmark datasets is at the cutting edge of the field, demonstrating its capability to understand and generate text based on visual content. This makes it a valuable tool for a wide range of applications in the domain of computer vision and natural language processing.


\begin{thebibliography}{00}
\bibitem{b1} Li, Junnan et al. “BLIP: Bootstrapping Language-Image Pre-training for Unified Vision-Language Understanding and Generation.” International Conference on Machine Learning (2022).
\bibitem{b2} Lin, Tsung-Yi ; Maire, Michael ; Belongie, Serge ; Bourdev, Lubomir ; Girshick, Ross ; Hays, James ; Perona, Pietro ; Ramanan, Deva ; Zitnick, C. Lawrence ; Dollár, Piotr: Microsoft COCO: Common Objects in Context, 2014
\bibitem{b3} Alane Suhr, , Stephanie Zhou, Ally Zhang, Iris Zhang, Huajun Bai, Yoav Artzi. "A Corpus for Reasoning About Natural Language Grounded in Photographs." (2019).
\bibitem{b4} Harsh Agrawal, , Karan Desai, Yufei Wang, Xinlei Chen, Rishabh Jain, Mark Johnson, Dhruv Batra, Devi Parikh, Stefan Lee, Peter Anderson. "nocaps: novel object captioning at scale." 2019 IEEE/CVF International Conference on Computer Vision (ICCV). IEEE, 2019.
\bibitem{b5}Zhe Gan, , Linjie Li, Chunyuan Li, Lĳuan Wang, Zicheng Liu, Jianfeng Gao. "Vision-Language Pre-training: Basics, Recent Advances, and Future Trends." (2022).
\bibitem{b6} Geoffrey Hinton, , Oriol Vinyals, Jeff Dean. "Distilling the Knowledge in a Neural Network." (2015).
\bibitem{b7} Suorong Yang, , Weikang Xiao, Mengcheng Zhang, Suhan Guo, Jian Zhao, Furao Shen. "Image Data Augmentation for Deep Learning: A Survey." (2022).
\bibitem{b8} Jacob Devlin, , Ming-Wei Chang, Kenton Lee, Kristina Toutanova. "BERT: Pre-training of Deep Bidirectional Transformers for Language Understanding." (2019).
\bibitem{b9} Alexey Dosovitskiy, , Lucas Beyer, Alexander Kolesnikov, Dirk Weissenborn, Xiaohua Zhai, Thomas Unterthiner, Mostafa Dehghani, Matthias Minderer, Georg Heigold, Sylvain Gelly, Jakob Uszkoreit, Neil Houlsby. "An Image is Worth 16x16 Words: Transformers for Image Recognition at Scale." (2021).
\bibitem{b10} Oriol Vinyals, Alexander Toshev, Samy Bengio, \& Dumitru Erhan. (2015). Show and Tell: A Neural Image Caption Generator.
\bibitem{b11} Ari Holtzman, Jan Buys, Li Du, Maxwell Forbes, \& Yejin Choi. (2020). The Curious Case of Neural Text Degeneration.
\bibitem{b12} Ranjay Krishna, Yuke Zhu, Oliver Groth, Justin Johnson, Kenji Hata, Joshua Kravitz, Stephanie Chen, Yannis Kalantidis, Li-Jia Li, David A. Shamma, Michael S. Bernstein, \& Fei-Fei Li. (2016). Visual Genome: Connecting Language and Vision Using Crowdsourced Dense Image Annotations.
\bibitem{b13} Adam Paszke, Sam Gross, Francisco Massa, Adam Lerer, James Bradbury, Gregory Chanan, Trevor Killeen, Zeming Lin, Natalia Gimelshein, Luca Antiga, Alban Desmaison, Andreas Köpf, Edward Yang, Zach DeVito, Martin Raison, Alykhan Tejani, Sasank Chilamkurthy, Benoit Steiner, Lu Fang, Junjie Bai, \& Soumith Chintala. (2019). PyTorch: An Imperative Style, High-Performance Deep Learning Library.
\bibitem{b14} Alec Radford, Jong Wook Kim, Chris Hallacy, Aditya Ramesh, Gabriel Goh, Sandhini Agarwal, Girish Sastry, Amanda Askell, Pamela Mishkin, Jack Clark, Gretchen Krueger, \& Ilya Sutskever. (2021). Learning Transferable Visual Models From Natural Language Supervision.


\end{thebibliography}
\end{document}